\documentclass{article}%
\usepackage{amsmath}
\usepackage{amsfonts}
\usepackage{amssymb}
\usepackage{graphicx}
\usepackage{rotating}
\usepackage[ruled,vlined,linesnumbered,noresetcount]{algorithm2e}
\usepackage{hyperref}
\usepackage{colortbl}
\usepackage{natbib}
\providecommand{\U}[1]{\protect\rule{.1in}{.1in}}
\newtheorem{theorem}{Theorem}

\newtheorem{lemma}[theorem]{Lemma}

\newtheorem{remark}[theorem]{Remark}

\newenvironment{proof}[1][Proof]{\noindent\textbf{#1.} }{\ \rule{0.5em}{0.5em}}
\begin{document}

\title{COVID-19 and other viruses: holding back its expansion by massive testing}
\author{Jos\'{e} Luis Sainz-Pardo$^{1}$, Jos\'{e} Valero$^{2}$\\{\small Centro de Investigaci\'{o}n Operativa, Universidad Miguel
Hern\'{a}ndez de Elche,}\\{\small Avda. Universidad s/n, 03202, Elche (Alicante), Spain}\\$^{1}${\small jose.sainz-pardo@umh.es}, $^{2}${\small jvalero@umh.es}\\}
\date{}
\maketitle

\begin{abstract}
The experience of Singapur and South Korea makes it clear that under certain
circumstances massive testing is an effective way for containing the advance
of the COVID-19.

In this paper, we propose a modified SEIR\ model which takes into account
tracing and massive testing, proving theoretically that more tracing and
testing implies a reduction of the total number of infected people in the long
run. We apply this model to the spread of the first wave of the disease in
Spain, obtaining numerical results.

After that, we introduce a heuristic approach in order to minimize the
COVID-19 spreading by planning effective test distributions among the
populations of a region over a period of time. As an application, the impact
of distributing tests among the counties of New York according to this method
is computed in terms of the number of saved infected individuals.

\end{abstract}

\bigskip


\bigskip


\section{Introduction}

The emergence of the coronavirus disease 2019 (COVID-19) and its mutations
motivated actions on the inhabitants of several countries like isolation,
social distance and others. The purpose of these restrictions is to slow down
the spread of the pandemic in order not to collapse health systems. Carrying
out massive testing is a complementary way to control the pandemic. Throughout
this article we will analyse the impact of testing on the number of infections
and we will explain how we have developed an expert system to obtain test
distributions which minimize the number of infections within a region and for
a temporal horizon.

The classical SIR model and its variations are a powerful mathematical tool
for predicting the evolution of epidemics around the world. The amount of
literature studying the properties of solutions of such models is huge, see
e.g. \cite{Brauer}, \cite{Jiao2020}, \cite{JiJiang}, \cite{KuniyaNakata},
\cite{LiMuldowney}, \cite{ZhangTeng}, \cite{ZhaoJiangRegan} and the references
therein among many others. In the present, much effort is taken in order to
estimate the evolution of the COVID-19 pandemic, see e.g. \cite{Annas},
\cite{Arcade}, \cite{Briton}, \cite{Chen}, \cite{ChenRui}, \cite{Gomes},
\cite{NdairouAreaNietoTorres}, \cite{Roda}, \cite{Sauter}, \cite{Xu}.

The coefficients of the models are constants in the simplest situation.
However, it is more realistic in order to estimate the evolution of an
epidemic to consider them as functions of time, as given in some of the above
references. Moreover, when we need to take into account the impact of
governmental actions like confinements, quarantines, restrictions on mobility
and travelling and so on, the coefficients, especially the rate of
transmission, can change abruptly. Thus, it is better to define them piecewise
by choosing appropriate time intervals depending on the moments in which the
governmental restrictions are more severe or, on the contrary, become more
relaxed. This approach has been considered in several papers estimating the
evolution of the COVID-19 pandemic in several countries, see \cite{Kwuimy},
\cite{Falco}, \cite{Lin}, \cite{Mushayabasa}, \cite{Niazi}, \cite{Tang}. Also,
the parameters can depend on time due to seasonality \cite{He}.

Apart from tough measures leading to a lockdown, tracing, massive testing and
isolation are very effective tools which help to contain the spread of the
illness even without restrictive social distancing, as we can see in the
examples of Singapur, South Korea and some Italian towns \cite{Romagnani}. The
main drawback of massive testing is the economic burden. However, as pointed
out in \cite{Eichenbaum} in the long run it pays off, as these policies can
dramatically reduce the economic costs of the epidemic.

Thus, it is quite interesting to modify the SIR or SEIR\ model in order to
measure their influence in the evolution of the pandemic and also to determine
the best way to carry out massive testing. In \cite{Berger} the SEIR\ model is
modified by taking into account testing and quarantine measures. It is shown
that increasing testing the governments could relax the quarantine conditions
(implementing a targeted and more efficient quarantine), so that the economic
and social costs would be smaller while maintaining constant the human costs.
In \cite{Wang} the author makes a qualitative analysis of a SIR\ model in
which the infected individuals are divided into two groups: those undetected
and those detected by means of some tests, which have a lower rate of
transmission of the disease due to quarantine measures. It is shown that
testing reduces the number of infections in the long term, avoiding in this
way herd immunity. In \cite{Niazi} a model taking into account random massive
tests is given. The impact of testing is analysed, obtaining the optimal
values for the number of tests to be carried out each day in two possible
testing policies to control the epidemic. In these papers, detection by
tracing and random testing are put together. In \cite{TangWang} a model in
which contact tracing and quarantine are considered is used in order to
estimate the evolution of the epidemic in Wuhan. In \cite{Ubaru} the problem
of\ optimal testing for individuals under limited testing and tracing
resources is studied by using a dynamic-graph based SEIR epidemiological model.

In this paper we modify the SEIR\ model in such a way that the impact of
tracing and random testing can be measured separately. For this aim we assume
that infected people which are detected are posed in quarantine and then they
cannot infect other people any more; hence, only undetected infected
individuals are able to spread the epidemic. In our model we differentiate two
kind of detected infected individuals: people which are detected because
either they are symptomatic or are direct contacts of infected individuals
(and then detected by tracing) and people which are detected by means of
massive random testing. These two methods of detection are complementary, as
by tracing one detects people with symptoms and their contacts, whereas by
testing one detects asymptomatic individuals that otherwise would remain
undetected and would continue infecting other susceptible individuals.

In Section \ref{Qualitative}, we make a qualitative analysis of the model in
the case where the coefficients are constants. We show first that, as in the
standard SEIR\ model, all the solutions converge as times increases to a fixed
point. The number of susceptible individuals in the long term determines the
size of the epidemic. We prove that the limit number of susceptible
individuals is an increasing function of the parameters of tracing and
testing. That is, the more tracing and testing, the less number of infected
people in the long term.

In Section \ref{Spain}, we estimate first the parameters of the model during
the first wave of the COVID-19 epidemic in Spain if no massive testing is
used. We consider the case where the parameters are functions of time and are
defined piecewise by taking into account the moments where confinement
restrictions were established or relaxed in Spain between March and June of
2020. After that we study the evolution of the epidemic when a constant number
of tests is carried out each day, calculating the number of infected
individuals which are saved in the long term for several values for the
parameters which characterize tracing and testing.

However, a constant distribution of tests per day is far to be optimal.
Moreover, it could be better to distribute the tests among different
populations in a non-proportional way. Thus, the main purpose of this work is
to develop an expert system that provides an effective distribution of tests
among the populations of a region over a period of time. So, in Section
\ref{distribution} a heuristic method based on the proposed modified
SEIR\ model is introduced in order to optimize the distribution of tests. In
this section it is also explained how to estimate the parameters of the model
by using the Differential Evolution technique. This technic was firstly
introduced in \cite{dev1} as an evolutionary method for optimizing nonlinear
functions. Also, it has been employed for estimating parameters of infectious
diseases models such as SIR, SIS, SEIR, SEIS and others \cite{dev2}. For
example, it has been applied to estimate the SIR parameters of the COVID-19
pandemic in Italy \cite{italia}. Several improvements, versions and
applications of this technique like \cite{dev4}, \cite{dev5} or the version
employed in \cite{dev3}, which will be the one used in this paper, can be
found in the literature.

Finally, in Section \ref{NY} we study the effectiveness of the proposed
heuristic approach by an extensive computational analysis of the spread of the
COVID-19 pandemic in the New York counties during the months of April, May and
June of 2020. The results of our distribution are also compared with a
distribution which is homogeneous in time and proportional to the size of each population.

\bigskip

\section{The model}

The classical SEIR\ model is the following%
\begin{equation}
\left\{
\begin{array}
[c]{c}%
\dfrac{dS}{dt}=-\dfrac{\beta}{N}SI,\\
\dfrac{dE}{dt}=\dfrac{\beta}{N}SI-\sigma E,\\
\dfrac{dI}{dt}=\sigma E-\gamma I,\\
\dfrac{dR}{dt}=\gamma I,
\end{array}
\right.  \label{SEIR}%
\end{equation}
where $N$ is the size of the population, $S\left(  t\right)  $ is the number
of the susceptible individuals to the disease, $E(t)$ is the number of exposed
people assuming that in the incubation period they do not infect anyone,
$I\left(  t\right)  $ is the number of currently infected individuals which
are able to infect other people, $R(t)$ is the number of individuals that have
been infected and then removed from the possibility of being infected again or
of spreading infection (which includes dead, recovered people and those in
quarantine or with immunity to the disease). The constant $\beta$ is the the
average number of contacts per person per time, $\gamma$ is the rate of
removal ($1/\gamma$ is the average time after which an infected individual is
removed), and $1/\sigma$ is the average time of incubation of the disease. All
these parameters are non-negative.

For our purposes we need to modify system (\ref{SEIR}) in several ways.

First, in a real situation the coefficients of the model are not constants but
functions of time. Moreover, these functions should not be continuous in
general, because in an epidemic outburst the governments impose restrictive
measures to the population leading to a sudden change of the rate of transmission.

Second, as our intention is to analyse the efficiency of a testing method, the
variable $I\left(  t\right)  $ will consist of all currently infected
individuals (not only of those able to infect)\ and a new variable $D\left(
t\right)  $, the number of currently infected people which are detected, will
be introduced. We will assume the ideal situation in which any detected
individual is placed in quarantine, so this person is not able to infect
anyone from that moment. Thus, the number of people with the capacity to
infect others is $I\left(  t\right)  -D\left(  t\right)  .$ Also, we do not
take into account that there could be people which are immune, so the variable
$R\left(  t\right)  $ will contain only dead and recovered individuals but
neither those in quarantine nor immune ones.

Third, we aim to estimate dead and recovered people among the detected ones
separately, so $R\left(  t\right)  $ is split into three variables:

\begin{itemize}
\item $F_{1}\left(  t\right)  $:\ number of dead individuals among the
detected ones;

\item $R_{1}\left(  t\right)  $:\ number of recovered individuals among the
detected ones;

\item $L\left(  t\right)  :$ number of removed individuals among the
undetected ones.
\end{itemize}

As a first step, we consider the situation where mainly people with symptoms
and their direct contacts are detected, but there is no a plan for massive
testing. The rate of detection is given by the variable $\rho\left(  t\right)
$ ($0<\rho\left(  t\right)  <1$) and, therefore, $D\left(  t\right)
=\rho(t)I(t).$

With these new variables at hand system (\ref{SEIR}) becomes:%
\begin{align}
\dfrac{dS}{dt}  &  =-\dfrac{\beta\left(  t\right)  }{N}S\left(  t\right)
(1-\rho\left(  t\right)  )I\left(  t\right)  ,\nonumber\\
\dfrac{dE}{dt}  &  =\dfrac{\beta\left(  t\right)  }{N}S\left(  t\right)
(1-\rho\left(  t\right)  )I\left(  t\right)  -\sigma E\left(  t\right)
,\nonumber\\
\dfrac{dI}{dt}  &  =\sigma E(t)-\rho\left(  t\right)  \left(  \gamma
_{1}(t)+\gamma_{2}(t)+(1-\rho\left(  t\right)  )\overline{\gamma}\left(
t\right)  \right)  I\left(  t\right)  ,\nonumber\\
\frac{dF_{1}}{dt}  &  =\gamma_{1}(t)\rho\left(  t\right)  I(t),\nonumber\\
\frac{dR_{1}}{dt}  &  =\gamma_{2}\left(  t\right)  \rho\left(  t\right)
I(t),\nonumber\\
\frac{dL}{dt}  &  =\overline{\gamma}(t)(1-\rho\left(  t\right)  )I(t),
\label{SEIR2}%
\end{align}
and the currently detected individuals are given by%
\begin{equation}
D(t)=\rho\left(  t\right)  I\left(  t\right)  . \label{Detected}%
\end{equation}
Here, $\gamma_{1}\left(  t\right)  $ is the rate of mortality of detected
people at moment $t$, $\gamma_{2}\left(  t\right)  $ stands for the rate of
recovery of detected people at moment $t$, whereas $\overline{\gamma}\left(
t\right)  $ is the rate of removal among those undetected.

In a second step, we describe the situation where a massive testing is planned
in order to increase the number of detected people, which can be placed then
in quarantine. For this aim we define the new variable $T\left(  t\right)  $,
which stand for the number of people that have been detected at time $t$ by
the test programme among those currently infected $I\left(  t\right)  $.

Hence, system (\ref{SEIR2}) becomes:%
\begin{align}
\dfrac{dS}{dt}  &  =-\frac{\beta\left(  t\right)  }{N}S\left(  t\right)
\left(  (1-\rho\left(  t\right)  )I\left(  t\right)  -T(t)\right)
,\nonumber\\
\frac{dE}{dt}  &  =\frac{\beta\left(  t\right)  }{N}S\left(  t\right)  \left(
(1-\rho\left(  t\right)  )I\left(  t\right)  -T(t)\right)  -\sigma E\left(
t\right)  ,\nonumber\\
\frac{dI}{dt}  &  =\sigma E(t)-\rho\left(  t\right)  \left(  \gamma
_{1}(t)+\gamma_{2}(t)+(1-\rho\left(  t\right)  )\overline{\gamma}\left(
t\right)  \right)  I\left(  t\right)  ,\nonumber\\
\frac{dT}{dt}  &  =\Delta(t)-\left(  \widetilde{\gamma}_{1}%
(t)+\widetilde{\gamma}_{2}(t)\right)  T(t),\nonumber\\
\frac{dF_{1}}{dt}  &  =\gamma_{1}(t)\rho\left(  t\right)
I(t)+\widetilde{\gamma}_{1}(t)T(t),\nonumber\\
\frac{dR_{1}}{dt}  &  =\gamma_{2}\left(  t\right)  \rho\left(  t\right)
I(t)+\widetilde{\gamma}_{2}(t)T(t),\nonumber\\
\frac{dL}{dt}  &  =\overline{\gamma}(t)(1-\rho\left(  t\right)
)I(t)-(\widetilde{\gamma}_{1}\left(  t\right)  +\widetilde{\gamma}_{2}\left(
t\right)  )T(t), \label{SEIR3}%
\end{align}
where $\Delta\left(  t\right)  $ is the number of detected people by testing
at day $t$ and $\widetilde{\gamma}_{1}(t)$ ($\widetilde{\gamma}_{2}(t)$) is
the rate of death (recovery) among those detected by testing. The currently
detected people are now calculated by%
\begin{equation}
D(t)=\rho\left(  t\right)  I\left(  t\right)  +T\left(  t\right)  .
\label{Detected2}%
\end{equation}
Assuming that the individuals for the testing are chosen randomly, the
variable $\Delta(t)$ is approximated by%
\[
\Delta(t)\simeq\alpha(t)\frac{(1-\rho(t))I(t)-T(t)}{N-R_{1}(t)-F_{1}\left(
t\right)  -\rho(t)I(t)-T(t)},
\]
with $\alpha(t)\geq0$ being the number of tests performed at day $t.$ This
formula can be simplified if we assume that $N>>R_{1}(t)+F_{1}\left(
t\right)  +\rho I(t)+T(t)$, which is true in big populations. Hence,
\begin{equation}
\Delta(t)\simeq\alpha(t)\frac{(1-\rho(t))I(t)-T(t)}{N}. \label{Delta}%
\end{equation}

For simplicity we could assume that the rates of death and recovery are the
same among the detected and the undetected infected people. In such a case,
$\overline{\gamma}(t)=\gamma_{1}\left(  t\right)  +\gamma_{2}\left(  t\right)
$, and $\widetilde{\gamma}_{1}\left(  t\right)  =\gamma_{1}\left(  t\right)
,\ \widetilde{\gamma}_{2}\left(  t\right)  =\gamma_{2}\left(  t\right)  $.
Thus, model (\ref{SEIR3})\ would be the following:%
\begin{align}
\dfrac{dS}{dt}  &  =-\frac{\beta\left(  t\right)  }{N}S\left(  t\right)
\left(  (1-\rho\left(  t\right)  )I\left(  t\right)  -T(t)\right)
,\nonumber\\
\frac{dE}{dt}  &  =\frac{\beta\left(  t\right)  }{N}S\left(  t\right)  \left(
(1-\rho\left(  t\right)  )I\left(  t\right)  -T(t)\right)  -\sigma E\left(
t\right)  ,\nonumber\\
\frac{dI}{dt}  &  =\sigma E(t)-\left(  \gamma_{1}(t)+\gamma_{2}(t)\right)
I\left(  t\right)  ,\nonumber\\
\frac{dT}{dt}  &  =\Delta(t)-\left(  \gamma_{1}(t)+\gamma_{2}(t)\right)
T(t),\nonumber\\
\frac{dF_{1}}{dt}  &  =\gamma_{1}(t)\left(  \rho\left(  t\right)
I(t)+T(t)\right)  ,\nonumber\\
\frac{dR_{1}}{dt}  &  =\gamma_{2}\left(  t\right)  \left(  \rho\left(
t\right)  I(t)+T(t)\right)  ,\nonumber\\
\frac{dL}{dt}  &  =(\gamma_{1}\left(  t\right)  +\gamma_{2}\left(  t\right)
)\left(  (1-\rho\left(  t\right)  )I(t)-T(t)\right)  . \label{SEIR4}%
\end{align}

\section{Qualitative analysis in the case of constant
coefficients\label{Qualitative}}

Our first aim is to prove theoretically that massive testing helps reducing
the number of infected people in the long-term.

The qualitative behaviour of the solutions of system (\ref{SEIR}) is simple
and well known \cite{Brauer}. There exists an interval of fixed points given
by%
\[
\left(  S_{\infty},E_{\infty},I_{\infty},R_{\infty}\right)  =\left(
S_{\infty},0,0,N-S_{\infty}\right)  ,\ S_{\infty}\in\lbrack0,N],
\]
and any solution with non-negative initial condition $\left(  S_{0}%
,E_{0},I_{0},R_{0}\right)  $ satisfying that $N=S_{0}+E_{0}+I_{0}+R_{0}$
converges to one of this fixed points as the time $t$ tends to $+\infty$.
Supposing that $R_{0}=0$, the limit point is determined by the initial value
of the susceptible $S$ and is given by the equation%
\begin{equation}
\log\frac{S_{0}}{S_{\infty}}=\frac{\beta}{\gamma}\left(  1-\frac{S_{\infty}%
}{N}\right)  . \label{FixedPointSEIR}%
\end{equation}

We consider system (\ref{SEIR4}) in the following particular situation:

\begin{itemize}
\item The coefficients $\beta,\ \rho,\ \sigma,\ \gamma_{1},\ \gamma
_{2},\ \widetilde{\gamma}_{1},\ \widetilde{\gamma}_{2}$, $\overline{\gamma}$
are constant.

\item The rates of death and recovery are the same among the detected and the
undetected infected people, so $\overline{\gamma}=\gamma=\gamma_{1}+\gamma
_{2},\ \widetilde{\gamma}_{1}=\gamma_{1},\ \widetilde{\gamma}_{2}=\gamma_{2}.$

\item The number of tests which are carried out per day $\alpha$ is constant
and positive and the approximation (\ref{Delta}) is valid.
\end{itemize}

With such assumptions and putting together the variables $F_{1},\ R_{1},\ L$
system (\ref{SEIR4}) reads as%
\begin{align}
\dfrac{dS}{dt}  &  =-\dfrac{\beta}{N}S((1-\rho)I-T),\nonumber\\
\dfrac{dE}{dt}  &  =\dfrac{\beta}{N}S((1-\rho)I-T)-\sigma E,\nonumber\\
\dfrac{dI}{dt}  &  =\sigma E-\gamma I,\nonumber\\
\dfrac{dT}{dt}  &  =\alpha\dfrac{\left(  1-\rho\right)  I-T}{N}-\gamma
T,\nonumber\\
\dfrac{dR}{dt}  &  =\gamma I. \label{SEIR5}%
\end{align}

What we want to show is that the solutions of system (\ref{SEIR5}) behave in
the same way as those of system (\ref{SEIR}) and that the limit value
$S_{\infty}$ is increasing with respect to the parameters $\alpha$ and $\rho$,
showing in this way theoretically that more detection implies a lower number
of infected people in the long-term.

The initial condition has to satisfy that $S_{0}+E_{0}+I_{0}+R_{0}=N$ and the
variable $W=S+E+I+R$ remains equal to $N$ for every time $t\geq0$. It is
straightforward to see that the unique fixed points of this system are:%
\[
\left(  S_{\infty},E_{\infty},I_{\infty},T_{\infty},R_{\infty}\right)
=\left(  S_{\infty},0,0,0,N-S_{\infty}\right)  ,\ S_{\infty}\in\lbrack0,N].
\]

We observe also that the initial condition has to be non-negative. This
implies that the solution is non-negative for every forward moment of time if,
moreover, $\left(  1-\rho\right)  I_{0}-T_{0}\geq0$.

\begin{lemma}
\label{Positive}If $S_{0},\ E_{0},\ I_{0},\ T_{0},\ R_{0}\geq0$ and $\left(
1-\rho\right)  I_{0}-T_{0}\geq0$, then $S\left(  t\right)  ,\ E\left(
t\right)  ,\ I\left(  t\right)  ,\ T\left(  t\right)  ,\ R\left(  t\right)
,\ y\left(  t\right)  =\left(  1-\rho\right)  I\left(  t\right)  -T\left(
t\right)  \geq0$ for all $t\geq0.$ Also, $S\left(  t\right)  >0$, for $t\geq
0$, if $S_{0}>0$, whereas $S\left(  t\right)  \equiv0$ if $S_{0}=0.$

Moreover, if the solution is not a fixed point, then $I\left(  t\right)
,\ T\left(  t\right)  ,\ R\left(  t\right)  >0$ for all $t>0$, and the
following statements hold true:

\begin{enumerate}
\item If either $E_{0}>0$ or $E_{0}=0$,\ $S_{0}>0$ and $y_{0}:=\left(
1-\rho\right)  I_{0}-T_{0}>0$, then $E\left(  t\right)  ,\ y\left(  t\right)
>0$ for all $t>0.$

\item If $E_{0}=0$, $S_{0}=0$ and $y_{0}>0$, then $E\left(  t\right)  =0$,
$y\left(  t\right)  >0$ for all $t>0.$

\item If $E_{0}=0$ and $y_{0}=0$, then $E\left(  t\right)  \equiv y\left(
t\right)  \equiv0.$
\end{enumerate}
\end{lemma}

\begin{proof}
It is easy to see from the first equation that $S\left(  t\right)  \geq0$ for
all $t\geq0.$ Moreover, $S\left(  t\right)  >0$, for all $t\geq0$, when
$S_{0}>0$ and $S\left(  t\right)  \equiv0$ if $S_{0}=0.$

If $E_{0}=I_{0}=0$, we have a fixed point. Thus, either $I_{0}>0$ or
$E_{0}>0.$

Let $E_{0}>0$. If $E\left(  t\right)  >0$ for all $t\geq0$, then it follows
from the third and fourth equations in (\ref{SEIR5}) that%
\begin{equation}
I(t)=I_{0}e^{-\gamma t}+\sigma\int_{0}^{t}e^{-\gamma\left(  t-r\right)
}E\left(  r\right)  dr>0, \label{I}%
\end{equation}%
\begin{equation}
T(t)=T_{0}e^{-\left(  \frac{\alpha}{N}+\gamma\right)  t}+\frac{\alpha\left(
1-\rho\right)  }{N}\int_{0}^{t}e^{-\left(  \frac{\alpha}{N}+\gamma\right)
\left(  t-r\right)  }I\left(  r\right)  dr>0, \label{T}%
\end{equation}
for all $t>0$, and the function $y\left(  t\right)  =\left(  1-\rho\right)
I\left(  t\right)  -T\left(  t\right)  $ satisfies%
\begin{equation}
\frac{dy}{dt}+\left(  \frac{\alpha}{N}+\gamma\right)  y=\left(  1-\rho\right)
\sigma E, \label{Eqy}%
\end{equation}
so that%
\begin{equation}
y\left(  t\right)  =y\left(  0\right)  e^{-\left(  \frac{\alpha}{N}%
+\gamma\right)  t}+\int_{0}^{t}e^{-\left(  \frac{\alpha}{N}+\gamma\right)
(t-r)}\left(  1-\rho\right)  \sigma E\left(  r\right)  dr>0,\ \forall t>0.
\label{y}%
\end{equation}
From the last equation in (\ref{SEIR5}) we infer that $R\left(  t\right)  >0$,
for every $t>0$, as well.

Next, we prove that the function $E\left(  t\right)  $ cannot vanish at any
point if $E_{0}>0$. Assume on the contrary the existence of a first moment of
time $t_{0}>0$ such that $E\left(  t_{0}\right)  =0$, so that $E\left(
t\right)  >0$ for all $t\in\lbrack0,t_{0})$. By the equality in (\ref{y}) we
conclude that $y\left(  t\right)  >0$ for $t\in\lbrack0,t_{0}]$. Hence, from
the second equation in (\ref{SEIR5}) we get%
\begin{equation}
E\left(  t_{0}\right)  =E_{0}e^{-\sigma t_{0}}+\dfrac{\beta}{N}\int_{0}%
^{t_{0}}e^{-\sigma(t_{0}-r)}S\left(  r\right)  ((1-\rho)I\left(  r\right)
-T\left(  r\right)  )dr>0, \label{E}%
\end{equation}
which is a contradiction.

Let now $I_{0}>0$ and $E_{0}=0$. Here we have to study three cases: 1)
$y_{0}>0,\ S_{0}>0;\ $2) $y_{0}>0,\ S_{0}=0$; 3) $y_{0}=0$.

Let $y_{0}>0$, $S_{0}>0$. By continuity there is $\varepsilon_{0}>0$ such that
$I\left(  t\right)  ,\ y\left(  t\right)  >0$ for $t\in\lbrack0,\varepsilon
_{0}]$. Making use of (\ref{T}) we obtain that $T\left(  t\right)  >0$ for all
$t\in(0,\varepsilon_{0}]$ as well. From the second equation in (\ref{SEIR5})
we have%
\[
E\left(  t\right)  =E_{0}e^{-\sigma t}+\frac{\beta}{N}\int_{0}^{t}%
e^{-\sigma(t-r)}S\left(  r\right)  y\left(  r\right)  dr>0\text{, for all
}t\in(0,\varepsilon_{0}],
\]
as $S\left(  t\right)  >0$ for every $t\geq0.$ Thus, taking any $0<\varepsilon
\leq\varepsilon_{0}$ we infer arguing as before but in the interval
$[\varepsilon,t]$ that $E\left(  t\right)  >0,$ for all $t\geq\varepsilon$,
and that $I\left(  t\right)  ,\ y\left(  t\right)  ,\ T\left(  t\right)
,\ R(t)>0,$ for all $t\geq\varepsilon$, as well. Since $\varepsilon$ is
arbitrarily small, $E\left(  t\right)  ,\ I\left(  t\right)  ,\ y\left(
t\right)  ,\ T\left(  t\right)  ,\ R\left(  t\right)  >0,$ for all $t>0.$

Let $y_{0}>0,\ S_{0}=0$. In this case, the first two equations in
(\ref{SEIR5}) imply that $S\left(  t\right)  =E\left(  t\right)  =0$ for all
$t\geq0$. Then from the third equation in (\ref{SEIR5}) and (\ref{Eqy})\ we
obtain that $I\left(  t\right)  ,\ y\left(  t\right)  >0$ for every $t\geq0$.
Finally, by (\ref{T}) and using the last equation in (\ref{SEIR5}) the
inequalities $T\left(  t\right)  ,\ R\left(  t\right)  >0,$ for all $t>0$,
hold true.

Let $y_{0}=0$. The fact that $\left(  E\left(  t\right)  ,y\left(  t\right)
\right)  \equiv\left(  0,0\right)  $ is the unique solution of the problem%
\[
\left\{
\begin{array}
[c]{c}%
\dfrac{dE}{dt}=\dfrac{\beta}{N}S(t)y-\sigma E,\\
\dfrac{dy}{dt}=-\left(  \dfrac{\alpha}{N}+\gamma\right)  y+\left(
1-\rho\right)  \sigma E,\\
E\left(  0\right)  =y\left(  0\right)  =0,
\end{array}
\right.
\]
implies that $y\left(  t\right)  \equiv0$ and $E\left(  t\right)  \equiv0$.
Then from the last three equations in (\ref{SEIR5}) and $T_{0}=\left(
1-\rho\right)  I_{0}>0$ we conclude that $I\left(  t\right)  ,$ $R\left(
t\right)  ,\ T\left(  t\right)  >0$ for $t>0.$
\end{proof}

\bigskip

\begin{theorem}
\label{Limit}Let $S_{0},\ E_{0},\ I_{0},\ T_{0},\ R_{0}\geq0$ and $\left(
1-\rho\right)  I_{0}-T_{0}\geq0$. Then every solution converges as
$t\rightarrow+\infty$ to one fixed point. Moreover, $S_{\infty}$ is determined
uniquely by $S_{0}$, $R_{0}$ and $T_{0}$ as the solution of the following
equation:%
\begin{equation}
\frac{\beta(1-\rho)}{\alpha+\gamma N}S_{\infty}-\log S_{\infty}=\frac
{\beta(1-\rho)}{\alpha+\gamma N}(N-R_{0})-T_{0}\frac{\beta}{\alpha+\gamma
N}-\log S_{0}. \label{EqSinf}%
\end{equation}

\end{theorem}

\begin{remark}
If the particular case where $\alpha=\rho=R_{0}=T_{0}=0$, we obtain
(\ref{FixedPointSEIR}).
\end{remark}

\begin{proof}
We divide the proof into several steps.

First, we state that $I\left(  t\right)  \rightarrow0,\ T\left(  t\right)
\rightarrow0$ as $t\rightarrow+\infty$.

Summing up the first three equations we have%
\begin{equation}
\frac{d}{dt}\left(  S+E+I\right)  =-\gamma I. \label{Eq1}%
\end{equation}
Then $W\left(  t\right)  =S\left(  t\right)  +E\left(  t\right)  +I\left(
t\right)  $ is a non-negative smooth non-increasing function, so it has a
limit as $t\rightarrow+\infty.$ Also, as
\[
W^{\prime\prime}\left(  t\right)  =-\gamma I^{\prime}\left(  t\right)
=-\gamma\left(  \sigma E\left(  t\right)  -\gamma I\left(  t\right)  \right)
\geq-\gamma\sigma N=-K,\ K\geq0,
\]
the derivative of $W$ tends to zero. Indeed, by contradiction assume that
there is $\nu>0$ and a sequence $t_{n}\rightarrow+\infty$ such that
$W^{\prime}\left(  t_{n}\right)  \leq-2\nu$. For $t\leq t_{n}$ it is clear
that%
\[
W^{\prime}\left(  t\right)  -W^{\prime}\left(  t_{n}\right)  =W^{\prime\prime
}\left(  \widetilde{t}\right)  \left(  t-t_{n}\right)  \leq-K\left(
t-t_{n}\right)  .
\]
Hence,%
\[
W^{\prime}\left(  t\right)  \leq-2\nu+K\left(  t_{n}-t\right)  ,
\]
so $W^{\prime}\left(  t\right)  \leq-\nu$ if $t\in J_{n}=[t_{n}-\frac{\nu}%
{K},t_{n}]$. The sequence $\{t_{n}\}$ can be chosen in such a way that the
intervals $J_{n}$ are disjoint. In each of them the function $W\left(
t\right)  $ decreases at least in the quantity $\frac{\nu^{2}}{K}$ and outside
them the function does not increase anywhere. Therefore,%
\[
W\left(  t_{n}\right)  \leq W\left(  0\right)  -n\frac{\nu^{2}}{K}\text{ for
any }n.
\]
For $n$ great enough the function becomes negative, which is not possible.
Thus, (\ref{Eq1}) implies that $I\left(  t\right)  \rightarrow0$ as
$t\rightarrow+\infty.$

As by Lemma \ref{Positive} we know that $0\leq T\left(  t\right)  \leq\left(
1-\rho\right)  I\left(  t\right)  $, it follows that $T\left(  t\right)
\rightarrow0$ as $t\rightarrow+\infty.$

Second, we establish that $E\left(  t\right)  \rightarrow0$ as $t\rightarrow
+\infty.$

We know that $W\left(  t\right)  =S\left(  t\right)  +E\left(  t\right)
+I\left(  t\right)  \rightarrow\eta$, $I\left(  t\right)  \rightarrow0$ as
$t\rightarrow+\infty.$ Also, as $S\left(  t\right)  $ is non-increasing and
bounded from below, $S\left(  t\right)  $ converges to some $S_{\infty}$.
Hence, $E\left(  t\right)  \rightarrow\eta-S_{\infty}$. We will prove that
$\eta-S_{\infty}=0.$ If not, there would exists $t_{0}>0$ such that%
\[
E\left(  t\right)  \geq\frac{\eta-S_{\infty}}{2}>0\text{ for all }t\geq
t_{0}.
\]
From the second equation in (\ref{SEIR5}) and $T\left(  t\right)  \geq0$,
$I\left(  t\right)  \rightarrow0$ we infer the existence of $t_{1}\geq t_{0}$
for which%
\[
\frac{dE}{dt}\leq-\sigma\frac{\eta-S_{\infty}}{4}\text{ for }t\geq t_{1}.
\]
Thus,%
\[
E\left(  t\right)  \leq E\left(  t_{1}\right)  -\sigma\frac{\eta-S_{\infty}%
}{4}\left(  t-t_{1}\right)  \rightarrow-\infty,
\]
which is not possible because $E\left(  t\right)  \geq0$.

The fact that every solution converges to a fixed point has been established.

Finally, let us prove (\ref{EqSinf}). From (\ref{Eq1}) and the above
convergences we have%
\begin{align*}
\int_{0}^{\infty}\left(  \frac{dS}{dt}+\frac{dE}{dt}+\frac{dI}{dt}\right)  dt
&  =-\gamma\int_{0}^{\infty}I\left(  t\right)  dt\\
&  =S_{\infty}-\left(  S_{0}+E_{0}+I_{0}\right)  =S_{\infty}-N+R_{0}.
\end{align*}
Integrating the fourth and first equations in (\ref{SEIR5}) and putting all
together we obtain%
\[
\int_{0}^{\infty}\frac{dT}{dt}dt=-T_{0}=\frac{\alpha\left(  1-\rho\right)
}{N}\int_{0}^{\infty}I\left(  t\right)  dt-\left(  \frac{\alpha}{N}%
+\gamma\right)  \int_{0}^{\infty}T\left(  t\right)  dt,
\]%
\begin{align*}
&  \log S_{\infty}-\log S_{0}\\
&  =-\frac{\beta\left(  1-\rho\right)  }{N}\int_{0}^{\infty}I\left(  t\right)
dt+\frac{\beta}{N}\int_{0}^{\infty}T\left(  t\right)  dt\\
&  =-\frac{\beta\left(  1-\rho\right)  }{N}\int_{0}^{\infty}I\left(  t\right)
dt+\frac{\beta\alpha\left(  1-\rho\right)  }{N(\alpha+\gamma N)}\int%
_{0}^{\infty}I\left(  t\right)  dt+T_{0}\frac{\beta}{\alpha+\gamma N}\\
&  =-\frac{\beta\left(  1-\rho\right)  }{N}\left(  1-\frac{\alpha}%
{\alpha+\gamma N}\right)  \left(  \frac{S_{\infty}-N+R_{0}}{-\gamma}\right)
+T_{0}\frac{\beta}{\alpha+\gamma N}\\
&  =-\frac{\beta\left(  1-\rho\right)  }{\alpha+\gamma N}\left(  N-S_{\infty
}-R_{0}\right)  +T_{0}\frac{\beta}{\alpha+\gamma N},
\end{align*}
giving rise to relation (\ref{EqSinf}).
\end{proof}

\bigskip

We aim now to analyse expression (\ref{EqSinf}) in order to show that
$S_{\infty}$ is an increasing function of $\alpha$ and $\rho$. We denote
$S_{\infty}\left(  \alpha\right)  $ the value of $S_{\infty}$ for $\alpha$
assuming the rest of the parameters being constant. In the same way we define
the function $S_{\infty}\left(  \rho\right)  .$

\begin{theorem}
We assume the conditions of Theorem \ref{Limit} and that $S_{0}>0$ and either
$E_{0}>0$ or $y_{0}>0$. Then $S_{\infty}\left(  \alpha_{1}\right)  >S_{\infty
}\left(  \alpha_{2}\right)  ,$ if $\alpha_{1}>\alpha_{2},$ and $S_{\infty
}\left(  \rho_{1}\right)  >S_{\infty}\left(  \rho_{2}\right)  ,$ if $\rho
_{1}>\rho_{2}.$
\end{theorem}

\begin{proof}
We write (\ref{EqSinf}) in the form%
\[
\left(  \alpha+\gamma N\right)  \log\frac{S_{\infty}}{S_{0}}+\beta\left(
1-\rho\right)  \left(  N-R_{0}-S_{\infty}\right)  -T_{0}\beta=0.
\]
We observe that Lemma \ref{Positive}\ implies that $y\left(  t\right)  >0$,
for $t>0$, so from the first equation in (\ref{SEIR5}) we obtain that
$S\left(  t\right)  $ is strictly decreasing and then $S_{\infty}<S_{0}$.
Analysing the function
\[
f\left(  x\right)  =\left(  \alpha+\gamma N\right)  \log\frac{x}{S_{0}}%
+\beta\left(  1-\rho\right)  \left(  N-R_{0}-x\right)  -T_{0}\beta
\]
we deduce that:

\begin{itemize}
\item $f\left(  S_{0}\right)  =\beta\left(  1-\rho\right)  \left(
N-R_{0}-S_{0}\right)  -T_{0}\beta=\beta\left(  \left(  1-\rho\right)
(I_{0}+E_{0})-T_{0}\right)  \geq0.$

\item $f$ has a maximum at $x_{0}=\dfrac{\alpha+\gamma N}{\beta\left(
1-\rho\right)  }$, $f\left(  x_{0}\right)  \geq0.$

\item $f$ is increasing if $0<x<x_{0}$ and $f\left(  x\right)  \rightarrow
-\infty$ as $x\rightarrow0^{+}.$

\item $f$ is decreasing if $x>x_{0}$ and $f\left(  x\right)  \rightarrow
-\infty$ as $x\rightarrow+\infty.$
\end{itemize}

We state that $f\left(  x_{0}\right)  >0.$ Indeed, if $f\left(  x_{0}\right)
=0$, then $x_{0}=S_{0}$ and this is moreover the only point where the function
vanishes. Then $S_{0}=S_{\infty}$, which is impossible because $S\left(
t\right)  $ is a strictly decreasing function.

Therefore, the above properties imply that the equation $f\left(  x\right)
=0$ has exactly two solutions: $0<x_{1}<x_{0},\ x_{2}>x_{0}.$ It follows from
$S_{\infty}<S_{0}$ and $S_{0}\in(x_{1},x_{2}]$ that $S_{\infty}=x_{1}$.

Denote by $g^{\alpha}\left(  x\right)  $ the function $f\left(  x\right)  $
for the value $\alpha$. It is easy to see that%
\[
g^{\alpha_{1}}\left(  x\right)  <g^{\alpha_{2}}\left(  x\right)  \text{ for
any }x<S_{0}\text{ if }\alpha_{1}>\alpha_{2},
\]
assuming that the other parameters are constant. From here we conclude that
$S_{\infty}\left(  \alpha_{1}\right)  >S_{\infty}\left(  \alpha_{2}\right)  $
if $\alpha_{1}>\alpha_{2}$.

In the same way, considering the functions $h^{\rho}\left(  x\right)  $ we
obtain that%
\[
h^{\rho_{1}}\left(  x\right)  <h^{\rho_{2}}\left(  x\right)  \text{ for any
}x\in(0,N-R_{0}]\text{ if }\rho_{1}>\rho_{2},
\]
while the other parameters remain constant. Hence, $S_{\infty}\left(  \rho
_{1}\right)  >S_{\infty}\left(  \rho_{2}\right)  $ if $\rho_{1}>\rho_{2}.$
\end{proof}

\bigskip

In the epidemic models a crucial parameter is the basic reproduction number,
which is the expected number of secondary cases produced, in a completely
susceptible population, by a typical infective individual over the course of
its infection period.

The variables $S$ and $R$ are disease free, whereas $E,I$ and $T$ are the
infective variables. We consider the equilibrium $\left(  N,0,0,0,0\right)  $
and the linearization around it of the subsystem of (\ref{SEIR5})
corresponding to the infective variables, which is given by%
\begin{equation}
\frac{dx}{dt}=\left(
\begin{array}
[c]{ccc}%
-\sigma & \beta\left(  1-\rho\right)  & -\beta\\
\sigma & -\gamma & 0\\
0 & \alpha\frac{1-\rho}{N} & -\frac{\alpha}{N}-\gamma
\end{array}
\right)  x=Jx. \label{Linearization}%
\end{equation}
Following \cite{Driessche} we calculate the reproduction number $R_{0}$ by
splitting the matrix $J$ into the rest of the Jacobian matrix associated to
the rate of new infections $F$ and the one associated to the net rate out of
the compartments $V$:%
\[
J=F-V=\left(
\begin{array}
[c]{ccc}%
0 & \beta\left(  1-\rho\right)  & -\beta\\
0 & 0 & 0\\
0 & \alpha\frac{1-\rho}{N} & -\frac{\alpha}{N}%
\end{array}
\right)  -\left(
\begin{array}
[c]{ccc}%
\sigma & 0 & 0\\
-\sigma & \gamma & 0\\
0 & 0 & \gamma
\end{array}
\right)  .
\]
The basic reproduction number is equal to the spectral radius of the matrix
$FV^{-1}$ when this matrix is non-negative. As the eigenvalues of the matrix%
\[
FV^{-1}=\left(
\begin{array}
[c]{ccc}%
\beta\frac{\left(  1-\rho\right)  }{\gamma} & \beta\frac{\left(
1-\rho\right)  }{\gamma} & -\frac{\beta}{\gamma}\\
0 & 0 & 0\\
\alpha\frac{1-\rho}{\gamma N} & \alpha\frac{1-\rho}{\gamma N} & -\frac{\alpha
}{\gamma N}%
\end{array}
\right)
\]
are $\lambda_{1}=0$ and $\lambda_{2}=\frac{1}{\gamma}\left(  \beta\left(
1-\rho\right)  -\frac{\alpha}{N}\right)  $, we obtain that
\[
R_{0}=\frac{1}{\gamma}\left(  \beta\left(  1-\rho\right)  -\frac{\alpha}%
{N}\right)
\]
provided that this quantity is non-negative, which is the usual situation as
$\alpha/N$ is small in great populations.

We can easily see that the zero solution of system (\ref{Linearization}) is
asymptotically stable if and only if $R_{0}<1$. Indeed, the characteristic
equation for the eigenvalues of the matrix $J$ is the following:%
\[
\lambda^{3}+a_{2}\lambda^{2}+a_{1}\lambda+a_{0}=0,
\]
where%
\begin{align*}
a_{2}  &  =\sigma+2\gamma+\frac{\alpha}{N},\\
a_{1}  &  =\left(  \sigma+\gamma\right)  \left(  \frac{\alpha}{N}%
+\gamma\right)  +\sigma\gamma,\\
a_{0}  &  =\sigma\gamma\left(  \frac{\alpha}{N}+\gamma-\beta\left(
1-\rho\right)  \right)  .
\end{align*}
According to the Routh-Hurwitz stability criterion, all the eigenvalues have
negative real part if and only if $a_{2},a_{0}>0$ and $a_{2}a_{1}-a_{0}>0$. It
is clear that%
\begin{align*}
a_{2}  &  >0,\\
a_{2}a_{1}-a_{0}  &  =\sigma\gamma\left(  \gamma+\sigma\right)  +\left(
\sigma+2\gamma+\frac{\alpha}{N}\right)  \left(  \sigma+\gamma\right)  \left(
\frac{\alpha}{N}+\gamma\right)  +\sigma\gamma\beta\left(  1-\rho\right)  >0,\\
a_{0}  &  =\sigma\gamma^{2}\left(  1-R_{0}\right)  .
\end{align*}
Thus, the result follows.

When the number of susceptible individuals is less than $N$, we replace
$R_{0}$ by the effective reproduction number at time $t$, given by%
\[
R_{t}=\frac{1}{\gamma}\left(  \frac{S\left(  t\right)  }{N}\beta\left(
1-\rho\right)  -\frac{\alpha}{N}\right)  .
\]

Finally, we observe that $R_{0}<1$ is equivalent to the inequality%
\[
\frac{\beta\left(  1-\rho\right)  }{\gamma+\frac{\alpha}{N}}<1.
\]
Thus, the coefficient $\widetilde{R}_{0}=\frac{\beta\left(  1-\rho\right)
}{\gamma+\frac{\alpha}{N}}$ could be also used as a reproduction number, which
corresponds to the one used in \cite{Niazi}.

\section{Validation of the model by applying it to the COVID-19 spread in
Spain\label{Spain}}

In this section, we estimate the parameters of model (\ref{SEIR2}) during the
first wave of the COVID-19 pandemic in Spain. After estimating the parameters,
we introduce the massive test detection. Solving system (\ref{SEIR3}) with a
constant number of test per day we show numerically that the final number of
susceptible $S_{\infty}$ would have been greater had a program for testing
been implemented. We show also the increasing dependence of $S_{\infty}$ with
resect to the parameter $\rho.$

We will assume that the rates of death and recovery are the same among the
detected and the undetected infected people. Therefore, $\overline{\gamma
}(t)=\gamma_{1}\left(  t\right)  +\gamma_{2}\left(  t\right)  $, and
$\widetilde{\gamma}_{1}\left(  t\right)  =\gamma_{1}\left(  t\right)
,\ \widetilde{\gamma}_{2}\left(  t\right)  =\gamma_{2}\left(  t\right)  $.

Taking into account that during the pandemic the government implemented at
certain moments of time restrictive measures of confinement leading to
reduction of mobility, following \cite{GutierrezVarona}, \cite{Lin},
\cite{Tang} the rate of transmission $\beta\left(  t\right)  $ will be a
piecewise continuous function with\ a finite number of discontinuities such
that in each interval of continuity the form of the function reads as:%
\[
\beta\left(  t\right)  =\beta_{0}-\beta_{1}\left(  1-e^{-\alpha\left(
t-t_{0}\right)  }\right)  .
\]
We extend this approach to the functions $\gamma_{1}\left(  t\right)
,\ \gamma_{2}\left(  t\right)  $, so that they are also piecewise defined
functions of the form:%
\[
\gamma_{i}(t)=\gamma_{0,i}-\gamma_{1,i}\left(  1-e^{-\alpha_{i}\left(
t-t_{0}\right)  }\right)  .
\]

As observed in \cite{Roda} the parameters $\rho\left(  t\right)  $ and
$\beta\left(  t\right)  $ are somehow dependent, which means that for a given
sample there exist several combinations of these parameters that fit well the
data. This is the problem of nonidentifiability. To avoid this drawback we use
the study of seroprevalence in Spain for the first wave of the epidemic and
choose an average value for the parameter $\rho$, so it is not estimated.

It has been estimated in \cite{Lauer} that the mean value of the incubation
period of the virus is about five days, so we take $\sigma=1/5$. There are
other studies which give a larger period of incubation. For example, in
\cite{Wu} the estimated value is around six.

Also, we choose an average value for $\rho$ given by the study of
seroprevalence in Spain \cite{Estudio}. According to this work, at the end of
May of 2020 5,2\% of the population of Spain had been infected by the virus
(which gives about 2400000 infected people as the population is 47 millions),
whereas an approximate number of 230000 people were detected by the COVID
tests at that moment. Thus, the average rate of detection during the first
wave of the pandemic in Spain was approximately equal to $0.1$.

We estimate the parameters of the model in the period from February 20, 2020
to May 17, 2020. Taking into account the points of confinement, we split this
interval into the following four subintervals: 1) 20/02-12/03; 2) 12/03-1/04;
3) 1/04-21/04; 4) 21/04-17/05.

We need to estimate the parameters of the functions $\beta\left(  t\right)
,\ \gamma_{1}\left(  t\right)  $ and $\gamma_{2}\left(  t\right)  $ in each
subinterval. For this aim we use the observed values of the variables
$D\left(  t\right)  ,\ F_{1}\left(  t\right)  ,\ R_{1}\left(  t\right)  $,
that is, the number of currently active infected, dead and recovered people
which were detected. We have taken the sample given by the Spanish Health
Ministry (see https://github.com/datadista/datasets/tree/master/COVID\%2019,
sections ccaa\_covid19\_confirmados\_pcr, ccaa\_covid19\_fallecidos,
ccaa\_covid19\_altas), using only the number of infected people detected by
means of a PCR\ test. The value of the constant $\alpha$ is $0$ and the
variable $T\left(  t\right)  $ is equal to $0$ as well (that is, there is no
massive testing). The observed value at time $t_{i}$ will be denote by
$D_{i},\ F_{1i}$ and $R_{1i}$, respectively. We consider in each interval the
pondered average of the euclidean norm of each observed variable:%
\begin{equation}
Error=\alpha_{1}\sqrt{\sum_{i=1}^{n}\left(  D_{i}-D\left(  t_{i}\right)
\right)  ^{2}}+\alpha_{2}\sqrt{\sum_{i=1}^{n}\left(  F_{1i}-F_{1}\left(
t_{i}\right)  \right)  ^{2}}+\alpha_{3}\sqrt{\sum_{i=1}^{n}\left(
R_{1i}-R_{1}\left(  t_{i}\right)  \right)  ^{2}}, \label{Error}%
\end{equation}
where $\alpha_{1}+\alpha_{2}+\alpha_{3}=1$. We have chosen $\alpha_{1}%
=\alpha_{2}=0.35,\ \alpha_{3}=0.3.$

On the 20th of February the number of detected active infected individuals was
$3$, so the estimate of the real number of infected subjects is $3/\rho=30$.
At that moment there were detected neither dead nor recovered people. We
assume that there were no removed subjects at all at the initial stage of the
pandemic. Hence, the initial value of the problem is given by:%
\[
I_{0}=30\text{, }F_{0}=0,\ R_{0}=0,\ L_{0}=0,\ S_{0}=N-I_{0}-E_{0}-F_{0}%
-R_{0}-L_{0}.
\]
As we do not have a hint for the value of $E_{0}$, we estimate it.

The estimate of the parameters is carried out by means of the minimization of
the target function (\ref{Error}) after solving system (\ref{SEIR2}) when the
values of the parameters go through a grid of points. The results in each
interval of time are the following:

\begin{enumerate}
\item 20/02-12/03: $\beta\left(  t\right)  =\beta_{0}=1.04,$ $\gamma
_{1}\left(  t\right)  =\gamma_{0,1}=0.0069,\ \gamma_{2}\left(  t\right)
=\gamma_{0,2}=0.014,$ $E_{0}=160.\ $In this interval, we have looked for
constant functions $\beta\left(  t\right)  ,\ \gamma_{1}\left(  t\right)
,\ \gamma_{2}\left(  t\right)  $, so $\beta_{1}=\gamma_{1,1}=\gamma_{1,2}=0$.

\item 12/03-1/04: $\beta\left(  t\right)  =0.6-0.596e^{-0.09\left(
t-21\right)  },$ $\gamma_{1}\left(  t\right)  =0.012-0.001e^{-0.05\left(
t-21\right)  },$ $\gamma_{2}\left(  t\right)  =0.016+0.04e^{-0.025(t-21)}.$

\item 1/04-21/04: $\beta\left(  t\right)  =0.04-0.033e^{-0.05(t-41)},$
$\gamma_{1}\left(  t\right)  =0.0095-0.008e^{-0.065(t-41)},$ $\gamma
_{2}\left(  t\right)  =0.055-0.025e^{-0.44(t-41)}.$

\item 21/04-17/05: $\beta\left(  t\right)  =0.02-0.0065e^{-0.09(t-61)},$
$\gamma_{1}(t)=0.0055-0.004e^{-0.075(t-61)},$ $\gamma_{2}\left(  t\right)
=0.025+0.01e^{-0.93(t-61)}.$
\end{enumerate}

\begin{figure}[th]
\centerline{\scalebox{0.6}{\includegraphics[angle=0]{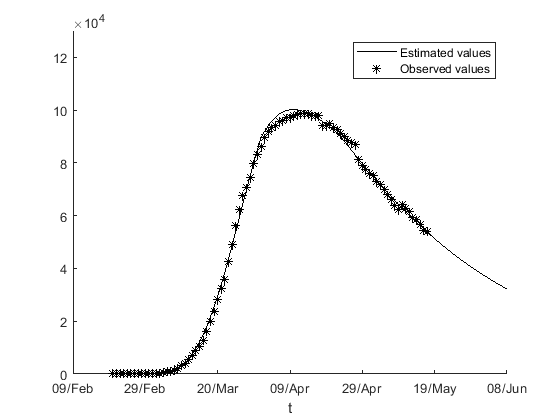}}}\caption{{\protect\footnotesize {Detected
currently infected individuals}}}%
\label{GrafInf}%
\end{figure}

\begin{figure}[th]
\centerline{\scalebox{0.6}{\includegraphics[angle=0]{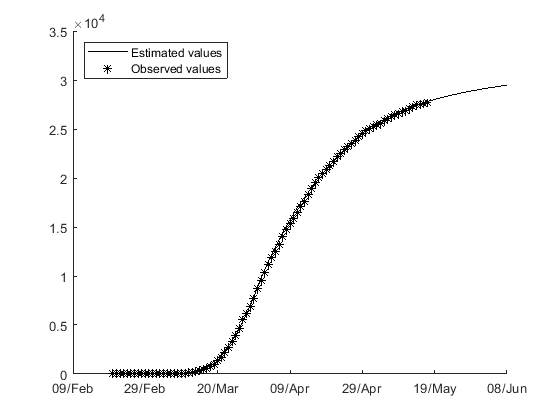}}}\caption{{\protect\footnotesize {Detected
dead individuals}}}%
\label{GrafDead}%
\end{figure}

\begin{figure}[th]
\centerline{\scalebox{0.6}{\includegraphics[angle=0]{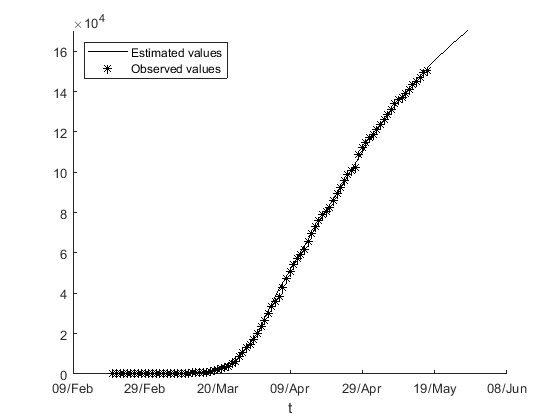}}}\caption{{\protect\footnotesize {Detected
recovered individuals}}}%
\label{GrafRecovered}%
\end{figure}

In figures \ref{GrafInf}-\ref{GrafRecovered} we can see the estimate of the
detected currently infected, dead and recovered individuals over the whole period.

\begin{figure}[th]
\centerline{\scalebox{0.6}{\includegraphics[angle=0]{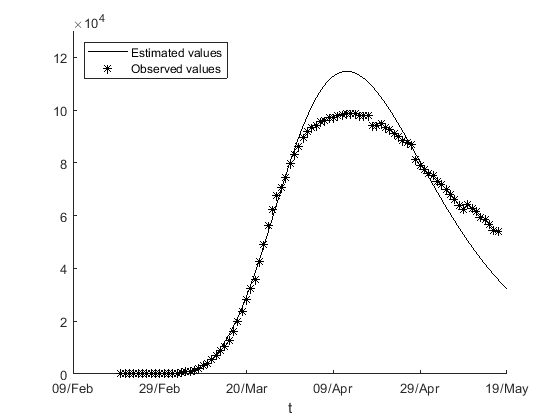}}}\caption{{\protect\footnotesize {Detected
currently infected individuals}}}%
\label{GrafInf2}%
\end{figure}

In figure \ref{GrafInf2} we can see the prediction given by the estimate of
the first two intervals, that is, using only the observed data until the first
of April. The number of currently infected people detected at the peak of the
pandemic is lower than the one given in the estimate. This reflects the fact
that on 28th of March the Spanish Government established more severe measures
of confinement by forbidding any non-essential activity. After some period
this restriction was withdrawn, so the slope of the curve of observed values
increased again.

Further we intend to estimate the impact of a massive random testing on the
final number of infected people during the first wave in Spain. For this aim
we solve system (\ref{SEIR4}) with three different values of the parameter
$\alpha$; namely, when we carry out $50000$, $100000$ and $150000$ random
tests per day.

\begin{figure}[th]
\centerline{\scalebox{0.6}{\includegraphics[angle=0]{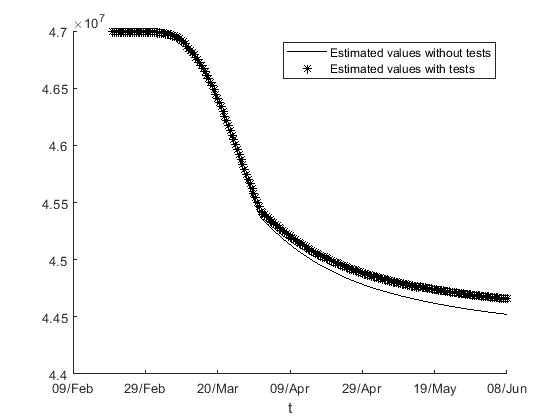}}}\caption{{\protect\footnotesize {Susceptible
individuals}}}%
\label{GrafSusceptible}%
\end{figure}

The approximate limit values of susceptible individuals $S_{\infty}$ in the
long run are the following:%
\[%
\begin{tabular}
[c]{|l|l|}\hline
$\alpha$ & $S_{\infty}$\\\hline
$0$ & $44364000$\\\hline
$50000$ & $44452000$\\\hline
$100000$ & $44535000$\\\hline
$150000$ & $44614000$\\\hline
\end{tabular}
\ \
\]
Therefore, the number of infections which are saved with massive testing is
$88000$, $171000$ and $250000$, respectively. In figure \ref{GrafSusceptible}
one can see the evolution of the number of susceptible without massive tests
and carrying out $100000$ tests per day.

We can measure also the effect of increasing the value of the parameter $\rho$
while the other parameters remain unchanged, and varying the value of $\alpha$
at the same time as well:%
\[%
\begin{tabular}
[c]{|l|l|l|l|}\hline
$\alpha$ & $\rho$ & $S_{\infty}$ & $%
\begin{array}
[c]{c}%
\text{number of saved}\\
\text{infections}%
\end{array}
$\\\hline
$0$ & $0.1$ & $44364000$ & \\\hline
$0$ & $0.15$ & $45372000$ & $1008\,000$\\\hline
$50000$ & $0.15$ & $45426000$ & $1062000$\\\hline
$100000$ & $0.15$ & $45476000$ & $1112\,000$\\\hline
$0$ & $0.3$ & $46592000$ & $2197\,000$\\\hline
$100000$ & $0.3$ & $46587000$ & $2223\,000$\\\hline
\end{tabular}
\
\]

It is important to observe that we have considered here the simple situation
where a constant number of random test is carried out each day. This procedure
is very far to be optimal. In the next section we will consider a heuristic
method to optimize the distribution of the tests in order to make massive
testing much more effective.

\section{Distribution approach\label{distribution}}

In the previous section we have analysed the impact of massive testing when
the number of tests which are carried out each day is constant. It is clear
that such an homogeneous distribution is not optimal. Therefore, in this
section we implement a heuristic method which allows us to increase the number
of saved infections. Unlike the previous situation, where we needed to
estimate the parameters of the model only once, now it will be necessary to
make the estimation a lot of times. Due to this, the parameters have to be
estimated using an automatic method. For this aim a genetic algorithm will be implemented.

\subsection{Estimation of the parameters}

The expert system that we have developed uses the Differential Evolution
technique for estimating the parameters of model (\ref{SEIR4}). The
differences between this method and other evolutionary algorithms are mainly
in the mutation and recombination phases in which weighted differences from
the space vector, despite random quantities, are used to obtain perturbations.
We have employed the Differential Evolution Algorithm described in
\cite{dev3}. Each individual is codified by an array containing the values of
the parameters to be estimated. The initial population is randomly generated
and Algorithm \ref{DE} describes the procedure which generates a new
population from the current population \textbf{P}$_{G}$ formed by $N$ individuals.

\begin{algorithm}[h]
	\setcounter{AlgoLine}{0}
	\label{DE}
	\caption{New\_Population(\textbf{P}$_G$)}
	\For{$i=1:N$}
	{
		Randomly select $r_1, r_2, r_3 \in \{1,2,...,N\}$ such that $r_1 \neq r_2 \neq r_3 \neq i$\\
		\textbf{u}$_{i,G+1} = $\textbf{x}$_{i,G} + K ($ \textbf{x}$_{r3,G}-$ \textbf{x}$_{i,G}) +
		F ($ \textbf{x}$_{r1,G} - $ \textbf{x}$_{r2,G}) $ \label{desc}\\
		\textbf{if} \textbf{u}$_{i,G+1}$ is better than \textbf{x}$_{i,G}$ \textbf{then} \textbf{x}$_{i,G+1}=$\textbf{u}$_{i,G+1}$ \textbf{else}  \textbf{x}$_{i,G+1}=$ \textbf{x}$_{i,G}$\\
	}
	\textbf{return} \textbf{P}$_{G+1}$
\end{algorithm}

Algorithm \ref{DE} generates one descendant for each individual $i$ belonging
to the current population \textbf{P}$_{G}$. Basically three different
individuals with index $r_{1}\neq r_{2}\neq r_{3}\neq i$ are randomly selected
from \textbf{P}$_{G}$. At Step \ref{desc}, the descendant \textbf{u}$_{i,G+1}$
is generated by \textbf{u}$_{i,G+1}=$\textbf{x}$_{i,G}+K($ \textbf{x}%
$_{r3,G}-$ \textbf{x}$_{i,G})+F($ \textbf{x}$_{r1,G}-$ \textbf{x}$_{r2,G})$.
The differential $K($ \textbf{x}$_{r3,G}-$ \textbf{x}$_{i,G})$ combines
\textbf{x}$_{r3,G}$ and \textbf{x}$_{i,G}$ whereas the differential $F($
\textbf{x}$_{r1,G}-$ \textbf{x}$_{r2,G})$ sets the step size. The coefficients
$K,F\in(0,1]$ are constants.

The selected individual is replaced by its descendant if it is better in the
sense that the value of a fitness function is lower. As in the previous
section, we use as the fitness function the pondered average of the euclidean
norm of each observed variable, given in (\ref{Error}). The algorithm is
repeated until no replacement happens after a maximal number of successive generations.

\subsection{Obtaining the planning distribution}

The aim of the distribution method is to plan the distribution of tests among
the locations of a region over the instants of time within a temporal horizon,
that is, to decide how many tests is better to assign to each location at each
instant of time. The objective of this planning is to minimize the total
number of infected individuals in a given region and in a temporal interval
or, what is the same, to maximize the number of saved infections. The main
purpose of such distribution is to plan an effective massive testing combined
with tracing.

The distribution method is based on forecasting the number of infected people
which would be saved after the assignment of one testing team to one location
at one instant of time within the temporal horizon. This procedure has to be
repeated for each location and each moment of time. Let ${\scriptsize {K}}$ be
the number of people tested per instant (day, week or month) by a testing
team, $L=\{l_{1},...,l_{e}\}$ be the set of locations (counties, towns, etc.),
and $T=\{1,2,...,M\}$ be the set of instants of time, being $t$ the current
instant. Firstly, we calculate the gain matrix $G=\{g_{l,t}\},$ that forecasts
the number of saved infections due to the assignment of ${\scriptsize {K}}$
tests at location $l$ and instant $t$. Secondly, the optimal test distribution
is obtained from this matrix according to a heuristic method that provides the
distribution matrix $D=\{d_{l,t}\},$ which indicates the number of tests to be
distributed in each location $l$ at each instant $t$. So, the procedures
$gain\_matrix$ (Algorithm \ref{st1}) and $test\_distribution$ (Algorithm
\ref{st2}) are used to obtain $G$ and $D$, respectively.

Algorithm \ref{st1} returns the gain matrix $G$. Initially, the parameters
($\beta(t)$, $\rho(t)$, $\sigma$, etc.) and the initial state variables
($S(t)$, $E(t)$, $I(t)$, etc.) are estimated by the Differential Evolution
technique for each location $l$ from the historical data of the coronavirus
spread until the current instant $t$. Then, for each location we solve system
(\ref{SEIR4}) twice with initial conditions at the moment $t$. Firstly, it is
solved with no test assignment. Secondly, it is solved but assigning
${\scriptsize {K}}$ tests only to the location $l$ and the instant $t+i$,
$1\leq i\leq14$. The difference of infected cases at $t+i+14$ in both
predictions provides the gain value $g_{l,t+i},$ that is, the number of saved
infections corresponding to testing ${\scriptsize {K}}$ people in the location
$l$ and instant $t$. However, if the effective reproduction number at time
$t+i$ and the location $l$, $R_{l,t+i}$, is lower than 1, it is considered
that there is no gain since the pandemic is in a decline phase, so
$g_{l,t+i}=0$. Besides, we observe that given that the circumstances are
variable and the expert system uses the procedure in a dynamic way, running it
every day, a myopic approach that focuses on obtaining gain values for the
next $14$ days after the current instant $t$ is used.

\begin{algorithm}[h]
	\setcounter{AlgoLine}{0}
	\label{st1}
	\caption{gain\_matrix}
	\ForEach{$l \in L$}
	{
		estimation of parameters and state variables for $l$ \label{reup}\\
	               \For{$i=1$ to $14$}
		{
			\If{$R_{l,t+i}\geq1$}
			{
				$\bar{p} = $prediction of infected cases at $t+14+i$ without testing\\
$p = $prediction of infected cases at $t+i+14$ assigning $\scriptsize{K}$ tests at $t+i$\\
				$g_{l,t+i}= \bar{p} - p$
			}
			\Else{$g_{l,t+i}= 0$}
		}				
	}
	\textbf{return} $G$
\end{algorithm}

Once the gain matrix is obtained, Algorithm \ref{st2} is called in order to
heuristically maximize the number of saved infections. So, first of all the
location-instant pair $(l^{\ast},t^{\ast})$ with largest gain value is chosen.
Then, as many tests as possible are allocated to the population $l^{\ast}$ and
instant $t^{\ast}>t$. This quantity $d_{l^{\ast},t^{\ast}}$ depends on several
constraints such as the number of remaining tests, the maximum number of
people which are able to be tested at this location per day, the number of
available test teams, etc. At Step \ref{subs}, the quantity of tests assigned
to $(l^{\ast},t^{\ast})$ is subtracted from the available tests. Then, the
procedure carries on locating the second largest gain, and so on. Once all the
available tests have been distributed the procedure ends up returning the
distribution matrix $D$.

\begin{algorithm}[h]
	\setcounter{AlgoLine}{0}
	\label{st2}
	\caption{test\_distribution}
	\While{$\#tests>0$}
	{
		$G_{l^*,t^*}$ = element of $G$ with maximum value\\
		$d_{l^*,t^*}$  = maximum number of feasible tests distributable to location $l^*$ and instant $t^*$\\
		$\#tests=\#tests-d_{l^*,t^*}$ \label{subs}\\
		$G_{l^*,t^*} = 0$\\
	}
	\textbf{return} $D$
\end{algorithm}

We note that it is convenient to refresh the estimate of parameters and the
state variables for each moment of time $t$, being necessary to compute again
both the gain and distribution matrices every time new data about the
coronavirus spread are reported. Thus, the proposed approach is used in a
dynamic way.

\section{Computational experience: distributing tests among the New York
counties\label{NY}}

In this section we will analyse the computational results of the proposed
distribution approach by measuring the number of saved infections and also by
comparing it with a distribution of tests which is homogeneous in time and
proportional to the size of each population. As an application, we will apply
this method to the New York state, which pandemic spread data are
disaggregated by counties and available in \cite{db}. The period of time
chosen for our study is from the first of April to the first of July of 2020.
The reason for choosing it is that it was during this period when the pandemic
spread was most virulent. Although the first case was detected on the first of
March, in the majority of the counties there were no registered cases until
the 15th of March. The historical data in this period are used in order to
simulate the pandemic spread and also to measure the effectiveness of the
possible distributions of tests. In this regard, the distribution obtained by
our approach is compared with the distribution of tests which is homogeneous
in time and proportional to the number of inhabitants of every county. It is
important to observe that as the data about the recovered individuals are not
reported, we have estimated them by supposing an average recovery time of $14$
days. Then, the historical recovery data have been built by using the
cumulative detected cases in the last 14 days for every moment of time.

Algorithm \ref{st3} shows the procedure that has been used in order to compute
the number of saved infections after applying our distribution method.
Initially, we estimate the parameters of model (\ref{SEIR4}) for the whole
period of study. However, as the conditions and parameters are different
according to different circumstances as the lockdown phases, the use of masks,
etc. the parameters are defined and estimated piecewise for each week. This is
similar to what is done in Section \ref{Spain} but now the intervals are fixed
in periods of seven days because the purpose of this study is to develop a
fully automatic decision system, avoiding thus the necessity to define the
intervals with dependence on the government restrictions as lockdown or
curfew. We will refer to these parameters as the \textit{simulation
parameters}.

Regarding the settings of the Differential Evolution technique which have been
used in this application, a population of $N=5$ individuals has been fixed.
Usually, the population size of genetic algorithms is higher but due to the
large number of parameters to be estimated this size has been adjusted to the
available RAM computer memory. The coefficients $K,F\in(0,1]$ were randomly
generated at each iteration. Finally, the estimation is stopped when a maximum
number of $1000$ iterations are carried out without any replacement . Our
numerical experiments were performed on a PC with a 2.33 GHz Intel Xeon dual
core processor, 8.5 GB of RAM, and with the operating system LINUX Ubuntu 18.04.5.

Secondly, for each day $t$, algorithms \ref{st1} and \ref{st2} are used for
planning the test distribution but computing the gain matrix using only the
historical data until $t$, given that we need to suppose the data of later
instants to be unknown. We need to define a matrix $D^{\prime}$ containing the
distribution of tests on each location at each moment of time over the whole
period. At any instant $t$ we only apply the obtained test distribution for
the next day, so the column $t+1$ in $D^{\prime}$ is replaced by the column
$t+1$ in the matrix $D$, that is $d_{l,t+1}^{\prime}=d_{l,t+1}$. We repeat
this procedure for each moment of time. We note that when we advance from $t$
to $t+1$ in order to obtain a new matrix distribution $D$ we need to estimate
again all the parameters, as the reported data is now available for one day
more. After computing the matrix $D^{\prime}$ we calculate the difference
between the estimated infected cases at time $M$ without testing $I_{0}=(M)$
and the estimated infected cases after applying our test distribution
$I_{D}^{\prime}(M),$ which provides the total number of saved infections.

\begin{algorithm}[h]
	\setcounter{AlgoLine}{0}
	\label{st3}
	\caption{saving}
	simulation parameters = estimated parameters until $M$\\
	$t=0$\\
	\While{$t<M$}
	{
		estimation of parameters = estimated parameters until t\\
		obtain $G$ and $D$ using the estimation parameters\\
		\textbf{for each} $l$ \textbf{do} $d'_{l,t+1}=d_{l,t+1}$\\		
	}
	compute $I_0(M)$ without testing and using the simulation parameters\\
	compute $I_{D'}(M)$ applying distribution $D'$ and using the simulation parameters\\
	\textbf{return} $I_0(M)-I_{D'}(M)$
\end{algorithm}

We note that in order to obtain the test distribution we employ the estimation
of parameters at each iteration, whereas the simulation parameters are used
for predicting the model and measuring the effectiveness of testing. Both
estimations do not need to have the same values since the first ones are
calculated with the information which is available until the current moment of
time and the simulation parameters are estimated using the data until the last
instant of time. Regarding the cases which are detected from the application
of random tests, they are estimated by using expression (\ref{Delta}), where,
for each location and instant $t$, $\alpha(t)$ is the number of tests to be
applied. Besides, in our computational experiments, the detected cases by
testing have been multiplied by a factor with the following values:

\begin{itemize}
\item 1, which means that the testing process is fully random and posteriorly
there is no COVID contact tracing;

\item 3, if it exists a process of COVID contact tracing which produces an
average of $2$ additional detected cases from the initially detected case;

\item 9, if besides the COVID contact tracing, testing is focused on towns or
groups of determined socio-demographic features, which allows to increase $3$
more times the probability of detection.
\end{itemize}

It is important to take into account that the product of this factor and the
number of daily tests $\alpha\left(  t\right)  $ cannot be greater that the
total population of the location.

We have carried out two types of computational experiments for different
quantities of available tests. In the first of them, the daily testing
capacity is restricted to $10000$ tests per day. In the second, the daily
testing capacity is restricted to $10\%$ of the total tests to apply in the
whole period.

Table \ref{table10000} shows the results obtained by our experiments. Column
$\#Tests$ indicates the number of tests to be assigned. These have been: 10,
50, 100 and 500 thousands. So, $12$ simulations, corresponding to the four
possible numbers of tests and the three factor values indicated in Column
$Factor$, have been carried out. \textit{Hom. Inf.} and \textit{Approach Inf.}
columns show the estimated infected cases by using the homogenous distribution
and our approach, respectively. On the other hand, \textit{Hom. Saving} and
\textit{Approach Saving} columns show the number of saved infections with both
methods. The number of infected cases from the first of April to the first of
June of 2020, including both those detected and not detected, have been
estimated in $3.365.817$. For this aim, model (\ref{SEIR4}) has been used,
estimating its parameters by Algorithm \ref{DE} with weekly intervals in which
the parameters can be different. No prevalence study was taking into account,
so $\rho(t)$ was also estimated for each interval. Then \textit{Hom. Saving}
and \textit{Approach Saving} were calculated as $3.365.817$ minus \textit{Hom.
Inf.} and \textit{Approach Inf.}, respectively. Finally, Column
\textit{Advantage} shows the difference \textit{Approach Saving minus Hom.
Saving}. Therefore, the greater the difference the more advantageous the
proposed approach is. On the contrary, negative values mean that the
homogeneous distribution overcomes our approach. We can see that the proposed
method improves the homogeneous distribution except in a few cases with
$500000$ tests. We note that such quantity of tests practically forces a
homogeneous distribution in time even with the proposed method. This explains
that the results are similar except when tracing and socio-demographic
features are taking into account, in which case the homogeneous method clearly
leads to better results. It is also important to observe that in this
particular case not all the available tests were distributed since $100000$
tests were not assigned.

\begin{table}[h]
\centering
\begin{tabular}
[c]{|r|r|r|r|r|r|r|}\hline
\# \textbf{Tests} & \textbf{Factor} & \textbf{Hom. Inf.} & \textbf{Approach
Inf.} & \textbf{Hom. Saving} & \textbf{Approach Saving} & \textbf{Advantage}%
\\\hline
10,000 & 1 & 3,365,384 & 3,365,783 & 34 & 433 & 399\\
10,000 & 3 & 3,364,519 & 3,365,716 & 101 & 1,298 & 1,197\\
10,000 & 9 & 3,361,923 & 3,365,514 & 303 & 3,894 & 3,591\\\hline
\rowcolor[gray]{0.8} 50,000 & 1 & 3,364,824 & 3,365,666 & 151 & 993 & 842\\
\rowcolor[gray]{0.8} 50,000 & 3 & 3,362,819 & 3,365,365 & 452 & 2,998 &
2,546\\
\rowcolor[gray]{0.8} 50,000 & 9 & 3,357,442 & 3,364,462 & 1,355 & 8,375 &
7,020\\\hline
100,000 & 1 & 3,364,744 & 3,365,519 & 298 & 1,073 & 775\\
100,000 & 3 & 3,362,612 & 3,364,924 & 893 & 3,205 & 2,312\\
100,000 & 9 & 3,357,124 & 3,363,144 & 2,673 & 8,693 & 6,020\\\hline
\rowcolor[gray]{0.8} 500,000 & 1 & 3,364,334 & 3,364,346 & 1,471 & 1,483 &
12\\
\rowcolor[gray]{0.8} 500,000 & 3 & 3,361,467 & 3,361,416 & 4,401 & 4,350 &
-51\\
\rowcolor[gray]{0.8} 500,000 & 9 & 3,354,169 & 3,352,700 & 13,117 & 11,648 &
-1,469\\\hline
\end{tabular}
\caption{Number of infected cases and saved infections with $10000$ tests per
day limitation}%
\label{table10000}%
\end{table}

Regarding the total number of tests which are employed, the more tests the
more infections are saved, but the ratio $saving/tests$ is decreasing as it
can be seen in Table \ref{ratio10000}, which shows the number of saved
infected cases per hundred applied tests. These ratios have been obtained for
both distributions with $factor=9$. On the one side, the effectiveness of the
proposed approach highly decreases with respect to test increments, whereas
the effectiveness of the homogeneous distribution is practically independent
of the total tests. On the other hand, the effectiveness of the proposed
distribution is notoriously higher but finally, due to the loss of its
effectiveness when increasing the number of tests, both are similar for
$500000$ tests. At last, Table \ref{calendario1} specifies the corresponding
distribution for $100000$ tests with $factor=1$.

\begin{table}[h]
\centering
\begin{tabular}
[c]{|r|r|r|}\hline
\textbf{\# Tests} & \textbf{$\%$ Homogeneous} & \textbf{$\%$ Approach}\\\hline
10,000 & 3.03 & 38.94\\\hline
50,000 & 2.71 & 16.75\\\hline
100,000 & 2.67 & 8.69\\\hline
500,000 & 2.62 & 2.33\\\hline
\end{tabular}
\caption{Ratios Saving/Tests with 10000 tests per day limitation}%
\label{ratio10000}%
\end{table}

\begin{table}[h]
\centering
\begin{tabular}
[c]{|l|l|r|}\hline
\textbf{Day} & \textbf{County} & \textbf{$\#$ Tests}\\\hline
12/04 & New York City & 10,000\\\hline
16/04 & Rensselaer & 10,000\\\hline
20/04 & Delaware & 10,000\\\hline
23/04 & Ulster & 10,000\\\hline
24/04 & Franklin & 10,000\\\hline
25/04 & Cortland & 10,000\\\hline
26/04 & Onondaga & 10,000\\\hline
27/04 & Fulton & 10,000\\\hline
28/04 & Oswego & 10,000\\\hline
29/04 & New York City & 10,000\\\hline
\end{tabular}
\caption{Planning for 100000 tests}%
\label{calendario1}%
\end{table}

Given that imposing a constant capacity of testing per day, which is
independent of the total number of tests, is a very strict limitation, we have
also carried out experiments in which the capacity per day was limited to
$10\%$ of the total number of tests. This computational experience is reported
in Table \ref{table10}. In this case the advantage of the proposed approach is
always higher than the homogeneous approach even in a new case with $1000000$
tests, which illustrates how high quantities of tests are also effective if
they are not highly restricted by the daily capacity of testing.

\begin{table}[h]
\centering
\begin{tabular}
[c]{|r|r|r|r|r|r|r|}\hline
\# \textbf{Tests} & \textbf{Factor} & \textbf{Hom. Inf.} & \textbf{Approach
Inf.} & \textbf{Hom. Saving} & \textbf{Approach Saving} & \textbf{Advantage}%
\\\hline
10,000 & 1 & 3,365,783 & 3,365,709 & 34 & 108 & 74\\
10,000 & 3 & 3,365,716 & 3,365,494 & 101 & 323 & 222\\
10,000 & 9 & 3,365,514 & 3,364,851 & 303 & 966 & 663\\\hline
\rowcolor[gray]{0.8} 50,000 & 1 & 3,365,666 & 3,365,280 & 151 & 537 & 386\\
\rowcolor[gray]{0.8} 50,000 & 3 & 3,365,365 & 3,364,209 & 452 & 1,608 &
1,156\\
\rowcolor[gray]{0.8} 50,000 & 9 & 3,364,462 & 3,361,026 & 1,355 & 4,791 &
3,436\\\hline
100,000 & 1 & 3,365,519 & 3,364,744 & 298 & 1,073 & 775\\
100,000 & 3 & 3,364,924 & 3,362,612 & 893 & 3,205 & 2,312\\
100,000 & 9 & 3,363,144 & 3,357,124 & 2,673 & 8,693 & 6,020\\\hline
\rowcolor[gray]{0.8} 500,000 & 1 & 3,364,346 & 3,360,256 & 1,471 & 5,561 &
4,090\\
\rowcolor[gray]{0.8} 500,000 & 3 & 3,361,416 & 3,354,512 & 4,401 & 11,305 &
6,904\\
\rowcolor[gray]{0.8} 500,000 & 9 & 3,352,700 & 3,341,480 & 13,117 & 24,337 &
1,1220\\\hline
1,000,000 & 1 & 3,362,884 & 3,355,417 & 2,933 & 10,400 & 7,467\\
1,000,000 & 3 & 3,357,058 & 3,344,645 & 8,759 & 21,172 & 12,413\\
1,000,000 & 9 & 3,339,880 & 3,316,234 & 25,937 & 49,583 & 23,646\\\hline
\end{tabular}
\caption{Number of infected cases and saved infections with $10\%$ tests per
day limitation}%
\label{table10}%
\end{table}

The $saving/test$ ratios of the distribution with the daily limitation of
$10\%$ of the total number of tests are reported in Table \ref{ratio10} for
$factor=9$. The cases with $10000$ and $50000$ tests show again how higher
restrictions of the daily capacity reduce effectiveness given that its daily
capacity is more restricted and as consequence its ratio decreases. The cases
with $500000$ and $1000000$ tests illustrate how a higher daily capacity
allows us to overcome the homogeneous distribution although the ratio
decreases. On the contrary, the ratio of the homogeneous distribution is less variable.

\begin{table}[h]
\centering
\begin{tabular}
[c]{|r|r|r|}\hline
\textbf{\# Tests} & \textbf{$\%$ Homogeneous} & \textbf{$\%$ Approach}\\\hline
10,000 & 3.03 & 9.66\\\hline
50,000 & 2.71 & 9.59\\\hline
100,000 & 2.67 & 8.69\\\hline
500,000 & 2.62 & 4.87\\\hline
1,000,000 & 2.60 & 4.96\\\hline
\end{tabular}
\caption{Ratios Saving/Tests with $10\%$ tests per day limitation}%
\label{ratio10}%
\end{table}

Table \ref{calendario2} specifies the corresponding distribution planning for
$1000000$ tests and $factor=1$. The counties of Chenango, Chemung, Delaware,
Franklin, Washington and Orleans are fully tested (the inhabitants for each
county have been obtained from \cite{population}). Note that for $100000$
tests the distribution planning is the same as the one reported in Table
\ref{calendario1}.

\begin{table}[h]
\centering
\begin{tabular}
[c]{|l|l|r|}\hline
\textbf{Day} & \textbf{County} & \textbf{$\#$ Tests}\\\hline
12/04 & Nassau & 100,000\\\hline
17/04 & Oswego & 100,000\\\hline
19/04 & Oneida & 100,000\\\hline
21/04 & Ontario & 100,000\\\hline
22/04 & Chemung & 88,830\\\hline
23/04 & Nassau & 100,000\\\hline
24/04 & Washington & 63,216\\\hline
25/04 & Franklin & 51,599\\\hline
26/04 & Delaware & 47,980\\\hline
26/04 & Orleans & 42,883\\\hline
27/04 & Chenango & 50,477\\\hline
28/04 & Westchester & 100,000\\\hline
29/04 & Rensselaer & 55,015\\\hline
\end{tabular}
\caption{Planning for 1000000 tests}%
\label{calendario2}%
\end{table}

\section{Conclusions}

In this work a SEIR model for analysing the efficiency of test distributions
has been introduced. It contemplates both detected and non-detected infected
individuals in order to measure the impact of testing. Since the values of the
parameters of the model can change abruptly due to severe governments measures
like lockdown, curfew, etc., the coefficients of the model are defined
piecewise in given intervals of time and are functions of time. This model has
been applied to the spread of the COVID\ pandemic in Spain. Besides, we have
theoretically proved how massive testing helps reducing the number of infected
people in the long-term.

Secondly, we describe the Differential Evolution technique, which is a genetic
algorithm for the estimation of parameters, and develop a heuristic approach
for distributing tests. This approach have been applied to the spread of
COVID\ pandemic in the New York counties by an extensive computational
experience showing the advantages of the proposed distribution method. Also,
an interesting future research line is to adapt a similar distribution
approach to the distribution of vaccines.

\bigskip

\textbf{Acknowledgements}

This work has been supported by the Generalitat Valenciana (Spain), project
2020/NAC/00022 . The first author has also been partially supported by the
Spanish Ministry of Science, Innovation and Universities, project
PGC2018-099428-B-I00. The second author has also been partially supported by
the Spanish Ministry of Science, Innovation and Universities, project
PGC2018-096540-B-I00, the Spanish Ministry of Science and Innovation, project
PID2019-108654GB-I00, and by Junta de Andaluc\'{\i}a (Spain) and FEDER,
projects P18-FR-2025 and P18-FR-4509.

\bibliographystyle{named}
\bibliography{testCovid}

\end{document}